\documentclass[letter,11pt]{article}
\usepackage{jheppub}
\usepackage{tikz}
\usetikzlibrary{decorations.pathmorphing}
\usetikzlibrary{decorations.markings}
\usepackage{xspace}

\usepackage{amsmath}
\usepackage{amssymb}
\usepackage{mathtools}
\usepackage[utf8]{inputenc}
\usepackage{physics}
\usepackage{slashed}
\usepackage{hyperref}
\usepackage[capitalize]{cleveref}

\newcommand{\ep}{\varepsilon}
\newcommand{\g}{\gamma}
\def\3g{{\gamma\gamma\gamma}}
\def\tr5{{\text{tr}_5}}

\title{Two-loop leading-colour QCD helicity amplitudes for two-photon plus jet production at the LHC}

\author[a]{Herschel A. Chawdhry,}
\author[b]{Micha\l{}  Czakon,}
\author[c]{Alexander Mitov,}
\author[c]{Rene Poncelet}

\affiliation[a]{Rudolf Peierls Centre for Theoretical Physics, Clarendon Laboratory, University of Oxford, Oxford OX1 3PU, United Kingdom}
\affiliation[b]{Institut f\"ur Theoretische Teilchenphysik und Kosmologie, RWTH Aachen University, D-52056 Aachen, Germany}
\affiliation[c]{Cavendish Laboratory, University of Cambridge, Cambridge CB3 0HE, United Kingdom}

\note{Preprint: OUTP-21-06P, Cavendish-HEP-21/05, P3H-21-014, TTK-21-09}

\abstract{We calculate the complete set of two-loop leading-colour QCD helicity amplitudes for $\gamma \gamma j$-production at hadron colliders. Our results are presented in a compact, fully-analytical form.}

\begin{document} 
\maketitle
\flushbottom

\section{Introduction}\label{sec:intro}

Multi-loop scattering amplitudes are core ingredients in high-precision perturbative calculations in Quantum Field Theory.
The complexity of these amplitudes rapidly increases with the number of loops, external legs, and kinematic scales in a process.
Scattering amplitudes in Quantum Chromodynamics (QCD) and Electroweak theory are of particular interest due to their central role in theoretical predictions for processes at the Large Hadron Collider (LHC).

The last few years have seen many advances in the calculation of multi-loop integrals and amplitudes~
\cite{Peraro:2016wsq,Chicherin:2018rpz,Kalin:2018thp,Dunbar:2019fcq,Caron-Huot:2019vjl,Dunbar:2020wdh,Anastasiou:2020sdt,Abreu:2020xvt,Dixon:2020bbt,
vonManteuffel:2014ixa,Chawdhry:2018awn,Kotikov:2018wxe,Bosma:2018mtf,Gehrmann:2018yef,Abreu:2018rcw,Boehm:2018fpv,Chicherin:2018mue,Mastrolia:2018uzb,Maierhofer:2018gpa,Kardos:2018uzy,Smirnov:2019qkx,Frellesvig:2019kgj,Bendle:2019csk,Papadopoulos:2019iam,Peraro:2019okx,Guan:2019bcx,Smirnov:2020quc,Usovitsch:2020jrk,Canko:2020ylt,Bendle:2020iim}.
Two-loop 5-point amplitudes are at the frontier of current amplitude calculations and have been the subject of particularly intense research~
\cite{Abreu:2018gii,Badger:2018gip,Abreu:2018jgq,Badger:2018enw,Abreu:2018zmy,Chicherin:2019xeg,Abreu:2019odu,Badger:2019djh,Hartanto:2019uvl,DeLaurentis:2020qle,Badger:2021nhg,Chawdhry:2019bji,Abreu:2020cwb,Chawdhry:2020for,Agarwal:2021grm,Abreu:2021fuk}.
That research has notably led to the calculation of the 2-loop QCD leading-colour amplitude for $q\bar{q} \rightarrow \3g$~\cite{Chawdhry:2019bji,Abreu:2020cwb,Chawdhry:2020for}, which in turn has enabled, for the first time, the computation of the Next-to-Next-to-Leading-Order (NNLO) QCD corrections for a $2\rightarrow 3$ process~\cite{Chawdhry:2019bji,Kallweit:2020gcp}.

In this work we calculate the complete set of 2-loop leading-colour QCD helicity amplitudes for the processes $q\bar{q}\rightarrow g\gamma\gamma$ and $qg\rightarrow q\gamma\gamma$. These amplitudes enable the calculation of the double-virtual QCD corrections to $\gamma \gamma j$ production at hadron colliders. They are also required, alongside the recently-calculated 3-loop QCD amplitude $q\bar{q} \rightarrow \gamma \gamma$~\cite{Caola:2020dfu}, to calculate the N\textsuperscript{3}LO QCD correction to $\gamma \gamma$ production, which is an important background for inclusive Higgs boson measurements at the LHC.

This paper is organised as follows. Our calculation is described in Section~\ref{sec:calculation}, including details of the infrared subtraction and helicity projections. Our results are presented in Section~\ref{sec:results} and our conclusions are presented in Section~\ref{sec:conclusion}. Some benchmark results are given in appendix~\ref{sec:benchmarks}. The analytic results derived in this paper are rather compact and are available for download in electronic form.

\section{Calculation}\label{sec:calculation}

We consider the partonic processes
\begin{eqnarray}
  q_c^{h_1}(p_1)\bar{q}^{h_2}_{c'}(p_2) \to g^{h_3}_a(p_3)\g^{h_4}(p_4)\g^{h_5}(p_5)\,,\nonumber\\
  q_c^{h_1}(p_1)g^{h_2}_{a}(p_2) \to q^{h_3}_{c'}(p_3)\g^{h_4}(p_4)\g^{h_5}(p_5)\,,
  \label{eq:process}
\end{eqnarray}
where $h_i \in \{+,-\}$ denotes the helicity of the $i$\textsuperscript{th} parton, $i=1,\dots , 5$. The indices $c$ and $c'$ denote the colours of the quarks while the index $a$ denotes the colour of the gluon.
The momenta $p_1$ and $p_2$ are incoming while $p_3$, $p_4$, and $p_5$ are outgoing.
All partons are massless and on-shell: $p_i^2 = 0$.
Momentum conservation and on-shell conditions leave five
independent parity-even Lorentz invariants $s_{ij} = (p_i+p_j)^2$ and one parity-odd invariant
$\tr5 = 4i\ep_{p_1p_2p_3p_4}$. We choose the following set of
variables to parametrise the amplitudes:
\begin{equation}
 x = \{s_{12},s_{23},s_{34},s_{45},s_{51},\tr5\} \;.
\label{eq:x} 
\end{equation}
All other Lorentz invariants can be expressed in terms of this set in the following way:
\begin{eqnarray}
  s_{13} &=& s_{12}-s_{23}-s_{45} \,,\\
  s_{14} &=& -s_{51}+s_{23}+s_{45}\,, \\
  s_{24} &=& s_{51}-s_{23}+s_{34} \,,\\
  s_{25} &=& s_{12}-s_{15}-s_{34} \,,\\
  s_{35} &=& s_{12}-s_{34}-s_{45}\;.
\end{eqnarray}
The physical scattering region satisfies \cite{Gehrmann:2018yef} the following equations:
\begin{equation}
  s_{12} > 0, ~~ s_{12} \geq s_{34}, ~~ s_{45}\leq s_{12} - s_{34}, ~~ s_{23} > s_{12}-s_{45},
  ~~ s_{51}^- \leq s_{51} \leq s_{51}^+ , ~~ (\tr5)^2 < 0\,,
\end{equation}
with
\begin{eqnarray}
 (\tr5)^2 &= &s_{12}^2(s_{23}-s_{51})^2 + (s_{23} s_{34} + s_{45} (s_{34} + s_{51}))^2- \nonumber\\
 & &2 s_{12} (s_{23}^2s_{34} + s_{23} s_{34} s_{45} - s_{23} (s_{34}+s_{45}) s_{51} + s_{45} s_{51} ( s_{34} + s_{51})) \,,
 \label{eq:tr5^2}
\end{eqnarray}
and
\begin{eqnarray}
 s_{51}^{\pm} &=& \frac{1}{(s_{12}-s_{45})^2}\biggl(s_{12}^2s_{23}+s_{12}s_{34}s_{45}-s_{23}s_{34}s_{45} -s_{34} s_{45}^2- \nonumber\\
  & &s_{12}s_{23}(s_{34}+s_{45}) \pm  2 \sqrt{s_{12} s_{23} s_{34} s_{45}(s_{45}+s_{23}-s_{12})(s_{45}+s_{34}-s_{12})} \biggr)\,.
\end{eqnarray}
The UV-renormalised amplitude for these
processes is denoted by:
\begin{equation}
 \mathcal{M}(\alpha_s)_{cc'a}^{f,h_1h_2h_3h_4h_5}(x) =
 \mathbf{T}^a_{cc'}
  \mathcal{M}(\alpha_s)^{f,h_1h_2h_3h_4h_5}(x)
   \equiv \mathbf{T}^a_{cc'} \mathcal{M}^{f,\bar{h}}\,,
\end{equation}
where we have factored out the tree-level colour structure. The index $f$ denotes the flavour structure, which is $f = q\bar{q}$ and $f = qg$ for the
$q\bar{q}$- and $qg$-initiated processes respectively. We summarise the
helicity configuration by $\bar{h} = \{h_1,h_2,h_3,h_4,h_5\}$
and suppress the kinematic dependence for brevity.
The amplitude can be expanded in $\alpha_s$:
\begin{equation}
 \mathcal{M}^{f,\bar{h}} = \sqrt{\alpha_s 4\pi}\left(\mathcal{M}^{f,\bar{h}(0)}
                    +\biggl(\frac{\alpha_s}{4\pi}\biggr)\mathcal{M}^{f,\bar{h}(1)}
                    +\biggl(\frac{\alpha_s}{4\pi}\biggr)^2\mathcal{M}^{f,\bar{h}(2)}
                    +\order{\alpha_s^3}\right)\;.
\label{eq:M-loop-expansion}                    
\end{equation}
The UV-renormalised amplitude $\mathcal{M}^{\bar{h}}$ is related to the bare amplitude computed in $d=4-2\ep$ dimensions $\mathcal{M}^{\bar{h},B}$ through:
\begin{equation}
 \mathcal{M}^{f,\bar{h}}(\alpha_s) =
   \left(\frac{\mu^2 e^{\gamma_E}}{4\pi}\right)^{-2\ep}
   \mathcal{M}^{f,\bar{h},B}(\alpha_s^0)\,.
\end{equation}
The bare coupling $\alpha_s^0$ is renormalised in the ${\overline{\rm MS}}$ scheme according to:
\begin{equation}
\alpha_s^0 = \left(\frac{e^{\gamma_E}}{4\pi}\right)^{\ep}
             \mu^{2\ep}Z_{\alpha_s} \alpha_s\,.
\end{equation}
The renormalisation constant $Z_{\alpha_s}$ is given in appendix \ref{sec:app-renorm}.

The IR divergences of the UV-renormalised amplitude can be factorised by means 
of the so-called $\mathbf{Z}$ operator: 
\begin{equation}
\mathcal{M}^{f,\bar{h}} = \mathbf{Z}^f \mathcal{F}^{f,\bar{h}}\,.
\label{eq:IR-fact}
\end{equation}
We define $\mathbf{Z}^f$ in the ${\rm \overline{MS}}$ scheme. Its explicit expression through 2 loops in QCD is given in appendix \ref{sec:app-renorm}. We note that eq.~(\ref{eq:IR-fact}) completely specifies the finite remainder $\mathcal{F}^{f,\bar{h}}$.

Once the $\mathbf{Z}^f$ factor, the finite remainder $\mathcal{F}$, and the amplitude $\mathcal{M}$ have been expanded in powers of $\alpha_s/(4\pi)$, eq.~(\ref{eq:IR-fact}) reduces to
\footnote{The $\alpha_s$ expansion of $\mathcal{F}^{f,\bar{h}}$ is analogous to the one in eq.~(\ref{eq:M-loop-expansion}).}
\begin{eqnarray}
  \mathcal{M}^{f,\bar{h},(0)} &=& \mathcal{F}^{f,\bar{h}(0)}\,,\\
  \mathcal{M}^{f,\bar{h},(1)} &=& \mathbf{Z}^{f,(1)}\mathcal{M}^{f,\bar{h}(0)}
                     +\mathcal{F}^{f,{h}(1)}\,,\\
  \mathcal{M}^{f,\bar{h},(2)} &=& \mathbf{Z}^{f,(2)}\mathcal{M}^{f,\bar{h}(0)}
                           +\mathbf{Z}^{f,(1)}\mathcal{F}^{f,\bar{h}(1)}
                           +\mathcal{F}^{f,\bar{h}(2)}\,.
\end{eqnarray}

The amplitude can be decomposed in terms of colour factors and the electric charges $Q_q$ and $Q_q'$ of, respectively, the external quark $q$ and the quarks $q'$ propagating in loops. The tree-level and 1-loop results will be given in sec.~\ref{sec:results} below. The complete 2-loop finite remainder, including the non-planar topologies,
can be decomposed as follows:
\begin{equation}
  \mathcal{F}^{(2)} = Q^2_q \mathcal{F}^{(2), Q^2_q}
   +Q_q Q_{l,1} \mathcal{F}^{(2), Q_qQ_{q'}}
   +Q_{l,2} \mathcal{F}^{(2), Q^2_{q'}}\,,
     \label{eq:F2-decomposition-Q}
\end{equation}
where we have introduced the abbreviation $Q_{l,n} = \sum_{q'} Q_{q'}^n$. 

In this work we only calculate the contributions from planar diagrams. To single them out, we expand the charge structures in the large-$N_c$ limit
\begin{eqnarray}
\mathcal{F}^{(2),Q^2_q}     &=& N_c^2\left(\mathcal{F}^{(2),Q^2_q,N_c^2}   + \frac{n_f}{N_c}\mathcal{F}^{(2),Q^2_q,n_f} + \order{{1\over N_c}}\right)\,, \nonumber\\
\mathcal{F}^{(2),Q_qQ_{q'}} &=& N_c  \left(\mathcal{F}^{(2),Q_qQ_{q'},N_c} + \order{{1\over N_c}}\right)\,,\nonumber\\
\mathcal{F}^{(2),Q^2_{q'}}  &=& N_c  \left(\mathcal{F}^{(2),Q^2_{q'},N_c}  + \frac{n_f}{N_c}\mathcal{F}^{(2),Q^2_{q'},n_f} + \order{{1\over N_c}}\right)\,,
\end{eqnarray}
where the $\order{1/N_c}$ terms receive contributions from non-planar diagrams. Furthermore, the contributions $\mathcal{F}^{(2),Q^2_{q'},N_c}$ and $\mathcal{F}^{(2),Q_qQ_{q'},N_c}$ also have contributing non-planar diagrams and are not considered in this work. The explicit expressions for the 2-loop finite remainders are given in sec.~\ref{sec:results} below.

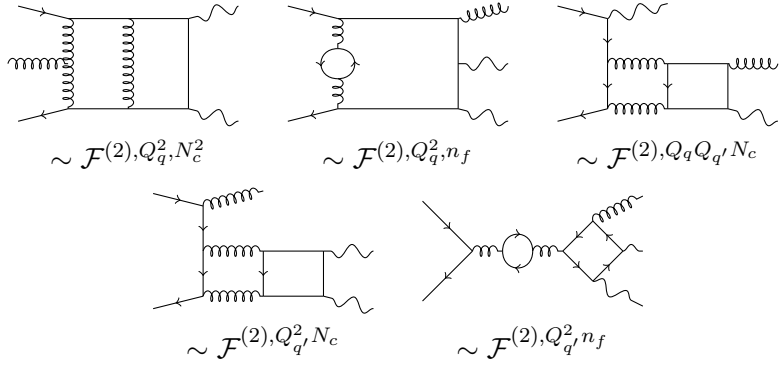
\begin{figure}[]
\centering

\tikzset{
    photon/.style={decorate, decoration={snake}},
    fermion/.style={postaction={decorate},
        decoration={markings,mark=at position .55 with {\arrow[]{>}}}},
    gluon/.style={decorate,
        decoration={coil,amplitude=2pt, segment length=3pt}} 
}

\begin{tikzpicture}[scale=0.4]

\node (caption) at (0,-3) { $\sim \mathcal{F}^{(2), Q^2_q,N_c^2}$  };
\node (dummy) at (0,-3.5) { };


%
\node (p1) at  (-4,2) {};
\node (p2) at  (-4,-2) {};
\node (p3) at  (4,-2) {};
\node (p4) at  (4,0) {};
\node (p5) at  (4,2) {};

\draw [fermion] (p1) to (-2,1.5) {};
\draw [fermion] (-2,-1.5) to (p2) {};
\draw [photon]  (p3) to (2,-1.5) {};
\draw [gluon] (-4,0) to (-2,0) {};
\draw [photon] (p5) to (2,1.5) {};

\draw  (-2,1.5) to (2,1.5) {};
\draw  (2,-1.5) to (-2,-1.5) {};
\draw  (2,1.5) to (2,-1.5) {};

\draw [gluon] (-2,1.5) to (-2,-1.5) {};
\draw [gluon] (0,1.5) to (0,-1.5) {};
\end{tikzpicture}
\begin{tikzpicture}[scale=0.4]

\node (caption) at (0,-3) { $\sim \mathcal{F}^{(2), Q^2_q,n_f}$  };
\node (dummy) at (0,-3.5) { };


%
\node (p1) at  (-4,2) {};
\node (p2) at  (-4,-2) {};
\node (p3) at  (4,-2) {};
\node (p4) at  (4,0) {};
\node (p5) at  (4,2) {};

\draw [fermion] (p1) to (-2,1.5) {};
\draw [fermion] (-2,-1.5) to (p2) {};
\draw [photon] (p3) to (2,-1.5) {};
\draw [photon] (p4) to (2,0) {};
\draw [gluon]  (p5) to (2,1.5) {};

\draw  (-2,1.5) to (2,1.5) {};
\draw  (2,-1.5) to (-2,-1.5) {};
\draw  (2,1.5) to (2,-1.5) {};

\draw [gluon] (-2,1.5) to (-2,0.5) {};
\draw [gluon] (-2,-0.5) to (-2,-1.5) {};
\draw [fermion] (-2,0.5) to[out=180,in=180,distance=0.75cm] (-2,-0.5) {};
\draw [fermion] (-2,-0.5) to[out=0,in=0,distance=0.75cm] (-2,0.5) {};
\end{tikzpicture}
\begin{tikzpicture}[scale=0.4]

\node (caption) at (0,-3) { $\sim \mathcal{F}^{(2), Q_qQ_{q'}N_c}$  };
\node (dummy) at (0,-3.5) { };


%
\node (p1) at  (-4,2) {};
\node (p2) at  (-4,-2) {};
\node (p3) at  (4,-2) {};
\node (p4) at  (4,0) {};
\node (p5) at  (4,2) {};

\draw [fermion] (p1) to (-2,1.5) {};
\draw [fermion] (-2,-1.5) to (p2) {};
\draw [photon] (p3) to (2,-1.5) {};
\draw [gluon]  (p4) to (2,0) {};
\draw [photon] (-2,1.5) to (0,2) {};

\draw [fermion] (-2,1.5) to (-2,0) {};
\draw [fermion] (-2,0) to (-2,-1.5) {};
\draw [gluon] (-2,0) to (0,0) {};
\draw [gluon] (-2,-1.5) to (0,-1.5) {};

\draw [fermion] (0,0) to (0,-1.5) {};
\draw  (0,0) to (2,0) {};
\draw  (2,0) to (2,-1.5) {};
\draw  (2,-1.5) to (0,-1.5) {};
\end{tikzpicture}

\begin{tikzpicture}[scale=0.4]

\node (caption) at (0,-3) { $\sim \mathcal{F}^{(2), Q_{q'}^2N_c}$  };
\node (dummy) at (0,-3.5) { };


%
\node (p1) at  (-4,2) {};
\node (p2) at  (-4,-2) {};
\node (p3) at  (4,-2) {};
\node (p4) at  (4,0) {};
\node (p5) at  (4,2) {};

\draw [fermion] (p1) to (-2,1.5) {};
\draw [fermion] (-2,-1.5) to (p2) {};
\draw [photon] (p3) to (2,-1.5) {};
\draw [photon]  (p4) to (2,0) {};
\draw [gluon] (-2,1.5) to (0,2) {};

\draw [fermion] (-2,1.5) to (-2,0) {};
\draw [fermion] (-2,0) to (-2,-1.5) {};
\draw [gluon] (-2,0) to (0,0) {};
\draw [gluon] (-2,-1.5) to (0,-1.5) {};

\draw [fermion] (0,0) to (0,-1.5) {};
\draw  (0,0) to (2,0) {};
\draw  (2,0) to (2,-1.5) {};
\draw  (2,-1.5) to (0,-1.5) {};
\end{tikzpicture}
\begin{tikzpicture}[scale=0.4]

\node (caption) at (0,-3) { $\sim \mathcal{F}^{(2), Q_{q'}^2n_f}$  };
\node (dummy) at (0,-3.5) { };


%
\node (p1) at  (-4,2) {};
\node (p2) at  (-4,-2) {};
\node (p3) at  (4,-2) {};
\node (p4) at  (4,0) {};
\node (p5) at  (4,2) {};

\draw [fermion] (p1) to (-2,0) {};
\draw [fermion] (-2,0) to (p2) {};
\draw [gluon] (-2,0) to (-1,0) {};
\draw [fermion] (-1,0) to[out=90,in=90,distance=0.75cm] (0,0) {};
\draw [fermion] (0,0) to[out=270,in=270,distance=0.75cm] (-1,0) {};
\draw [gluon] (0,0) to (1,0) {};
\draw [fermion] (1,0) to (2,-1) {};
\draw [fermion] (2,-1) to (3,0) {};
\draw [fermion] (3,0) to (2,1) {};
\draw [fermion] (2,1) to (1,0) {};

\draw [photon] (3,0) to (p4) {};
\draw [photon] (2,-1) to (p3) {};
\draw [gluon]  (2,1) to (p5) {};
\end{tikzpicture}
\caption{Representative two-loop diagrams for the contributing colour/charge coefficients.
}\label{fig-diagrams}
\end{figure}

The partonic process $q\bar{q} \to g \gamma\gamma$ given in eq.~(\ref{eq:process}) has three independent helicities. We choose the following set:
\begin{equation}
  \mathcal{F}^{q\bar{q},\{+----\}} \quad , \quad \mathcal{F}^{q\bar{q},\{+-+--\}} \quad \text{and} \quad \mathcal{F}^{q\bar{q},\{+--+-\}}\,.
  \label{eq: helicities-qqbar}
\end{equation}
All other helicity configurations can be obtained from the above three by conjugation and/or permutation of the external momenta. We note that the helicity amplitudes $\mathcal{F}^{q\bar{q},\{+----\}}$, $\mathcal{F}^{q\bar{q},\{+++--\}}$ and $\mathcal{F}^{q\bar{q},\{+-+++\}}$ vanish at tree-level. For this reason they do not contribute to the 2-loop squared matrix element and enter NNLO QCD computations only through the 1-loop-squared matrix element.

The partonic process $qg\to q\gamma\gamma$ specified in eq.~(\ref{eq:process}) can similarly be expressed in terms of three independent helicities, and the remaining helicities can be obtained by conjugation and/or permutation of external legs. Unlike the $q\bar q$ initiated process, however, the $qg$ one involves crossings between the initial and the final states. Such crossings involve a practical complication related to the nature of the pentagon functions \cite{Chicherin:2020oor} used to express these amplitudes; it originates in the fact that these functions' representation does not allow crossings between the initial and final states in the course of their numerical evaluation. For this reason, we have computed analytically six helicities from which all remaining ones can be obtained by crossing final-state momenta only: 
\begin{eqnarray}
  && \mathcal{F}^{qg,\{+-+--\}} \quad , \quad \mathcal{F}^{qg,\{++++-\}} \quad \text{and} \quad \mathcal{F}^{qg,\{+++--\}}\,,\nonumber\\
  && \mathcal{F}^{qg,\{+-++-\}} \quad , \quad \mathcal{F}^{qg,\{+++++\}} \quad \text{and} \quad \mathcal{F}^{qg,\{+-+++\}}\,.
  \label{eq: helicities-qg}
\end{eqnarray}
All other helicities for this process can be obtained from the above six by conjugation and/or permutation of final-state momenta. 

Just as in our recent 3-photon calculation~\cite{Chawdhry:2020for}, we have decomposed the amplitudes using the helicity projection method proposed in ref.~\cite{Chen:2019wyb}. The construction of the gluon/photon polarization vector follows closely ref.~\cite{Chawdhry:2020for}. The fermion projector in our previous work was designed with the permutation symmetry of the final-state photons in mind. The partonic processes considered here are less symmetric and thus we consider a simpler projector. In the case of the $q\bar{q} \to g \gamma\gamma$ partonic process, the fermionic line is projected by considering the following trace in spin space:
\begin{equation}
  \mathcal{M} = \bar{v}(p_2,h_2) \Gamma u(p_1,h_1)
              = \Tr \biggl\{ \left(u \otimes \bar{v}\right)
                  \Gamma \biggr\}\,,
\end{equation}
where $\Gamma$ represents the spinor-stripped amplitude. The density operator $u\otimes\bar{v}$ can be rearranged to
\begin{equation}
  (u\otimes\bar{v})_{\alpha\beta}= \frac{1}{\bar{u}Nv}[(u\otimes\bar{u}) N
              (v \otimes \bar{v})]_{\alpha\beta}\,,
  \label{eq:u-times-vbar}
\end{equation}
with any $N$ for which $\bar{u}Nv \neq 0$. We choose here $N = \slashed{p_3}$. The projectors for the $qg \to q \gamma\gamma$ process are obtained from those for $q\bar{q} \to g \gamma\gamma$ by applying the crossing $p_2 \leftrightarrow p_3$.

The amplitudes are expressed in terms of a minimal basis of irrational functions by means of the same automated framework that we used in the 3-photon amplitude calculation \cite{Chawdhry:2020for}.
Our framework performs finite-field sampling of the rational coefficients appearing in the amplitude by evaluating and combining the analytical IBP solutions from ref.~\cite{Chawdhry:2018awn} and the master integral solutions from ref.~\cite{Chicherin:2020oor}.
The analytical results for the amplitude are obtained by interpolating the finite-field samples using the library {\tt FireFly}~\cite{Klappert:2020aqs}.
We refer the reader to ref.~\cite{Chawdhry:2020for} for a detailed description of our framework.

\section{Results}\label{sec:results}

We write the finite remainder, suppressing the process ($f$) and helicity ($\bar{h}$) indices, in the following way:
\begin{eqnarray}
 \mathcal{F} &=& Q^2_q \mathcal{F}^{(0),Q^2_q} \biggl(
    1+ \biggl(\frac{\alpha_s}{4\pi}\biggr)\biggl(
     C_F \mathcal{R}^{(1),Q_q^2,C_F}+\frac{T_F}{C_A} \mathcal{R}^{(1),Q_q^2,T_F/C_A}
 + T_F \frac{Q_{l,2}}{Q^2_q}\mathcal{R}^{(1),Q_{q'}^2,T_F} \biggr)\nonumber\\
  & &+ \biggl(\frac{\alpha_s}{4\pi}\biggr)^2\biggl(
     N_c^2 \mathcal{R}^{(2),Q_q^2,N_c^2}+ N_c n_f \mathcal{R}^{(2),Q_q^2,n_f}
               +n_f \frac{Q_{l,2}}{Q^2_q}\mathcal{R}^{(2),Q_{q'}^2,n_f}\biggr)\biggr)\;.
\label{eq:F-R}
\end{eqnarray}
In case of a vanishing tree-level amplitude we write 
\begin{eqnarray}
 \mathcal{F} &=& Q^2_q \tilde{\mathcal{F}}^{(0),Q^2_q} \biggl(
    \biggl(\frac{\alpha_s}{4\pi}\biggr)\biggl(
     C_F \mathcal{R}^{(1),Q_q^2,C_F}+\frac{T_F}{C_A} \mathcal{R}^{(1),Q_q^2,T_F/C_A}
 + T_F \frac{Q_{l,2}}{Q^2_q}\mathcal{R}^{(1),Q_{q'}^2,T_F} \biggr)\nonumber\\
  & &+ \biggl(\frac{\alpha_s}{4\pi}\biggr)^2\biggl(
     N_c^2 \mathcal{R}^{(2),Q_q^2,N_c^2}+ N_c n_f \mathcal{R}^{(2),Q_q^2,n_f}
               +n_f \frac{Q_{l,2}}{Q^2_q}\mathcal{R}^{(2),Q_{q'}^2,n_f}\biggr)\biggr)\;.
\label{eq:F-R-tilde}
\end{eqnarray}

Our fully-analytic results for the remainders $\mathcal{R}^{(\ell),i,c}$, with $\ell=1,2$ and $i$/$c$ labelling the charge/colour structure, and for the normalisations $\mathcal{F}^{(0),Q^2_q}$ and $\tilde{\mathcal{F}}^{(0),Q^2_q}$ can be found in the supplementary material attached to this paper. 

The remainders $\mathcal{R}^{(\ell),i,c}$ have the following structure:
\begin{equation}
 \mathcal{R}^{(\ell),i,c} = \sum_e r^{(\ell),i,c}_e\; t_e\,.
 \label{eq:R}
\end{equation}
The coefficients $r^{(\ell),i,c}$ are rational functions of $s_{ij}$ and linear functions of $\tr5$. By $t_e$ we denote the elements of the transcendental basis. The parity-odd variable $\tr5$ has two distinct origins: the helicity projection and the master integrals of ref.~\cite{Chicherin:2020oor}. Scalar integrals only depend on the invariants $s_{ij}$, however a physical phase space needs additionally the sign of $\Im{\tr5}$ to be fully specified and this dependence is kept explicit in this master integral representation. Internally, we keep these two separate. We relabel the $\tr5$ originating from the masters as $\ep_5$ and include it as part of the basis $t_e$. Since $\ep_5$ only appears together with parity-odd combinations of pentagon functions, absorbing $\ep_5$ in the transcendental basis effectively renders all elements $t_e$ parity-even.

A comment is in order about the crossing of momenta in the presence of the parity-odd variable $\ep_5$ and the parity-odd transcendental functions appearing in the basis of ref.~\cite{Chicherin:2020oor}. When the analytic expression for a given helicity is being derived, one needs to match scalar integrals with permuted momenta to the basis $t_e$. In practice, these scalar integrals are the master integrals resulting from the IBP reduction of the amplitude. 
This matching is done in the following way. Ref.~\cite{Chicherin:2020oor} provides the transcendental basis $t_e$ indirectly, through a set of master integrals of uniform transcendentality (UTM). These UTMs depend on the invariants $s_{ij}$ but not on $\ep_5$. Ref.~\cite{Chicherin:2020oor} also provides for a single permutation of momenta a way of expressing a set of scalar integrals to the set of UTMs. The parity-odd variable $\ep_5$ appears linearly in this relation. This relation allows us to map our own scalar integrals to UTMs and, from there, to $t_e$. 

Additionally, ref.~\cite{Chicherin:2020oor} provides for each UTM its complete set of permutations in all five momenta. In practice this means that we do not need to perform any crossing of momenta in the transcendental functions appearing in the basis $t_e$, but simply need to identify the UTMs with the correct crossing that matches the crossing of the scalar integral. The variable $\ep_5$, however, needs to be treated separately under permutation of momenta since it is not part of the pre-crossed set of UTMs. In particular, its sign changes under odd permutations. 

When one numerically crosses momenta in a given helicity amplitude (for example, in order to derive a different helicity from a known one) the above procedure needs to be slightly modified. Since in this case one cannot rely on the set of pre-crossed UTMs anymore, one has to treat correctly the signs of both the parity-odd variable $\ep_5$ and the parity-odd functions within the transcendental basis $t_e$. Specifically, if an odd permutation is involved the signs of both flip. As we previously remarked, the transcendental functions of ref.~\cite{Chicherin:2020oor} allow for the numerical evaluation of crossings only if the momenta are not switched between the initial and the final states.

We have simplified the rational functions $r^{(\ell),i,c}$ appearing in the amplitudes by expressing them as linear combinations of a much smaller set of rational functions. Those have been further simplified with the help of the partial fractioning package {\tt MultivariateApart}~\cite{Heller:2021qkz}. In the following we will be referring to this much smaller set as {\it independent} rational functions. The number of (independent) rational functions for each colour/charge structure can be found in table~\ref{table:rational}. The appearance of these simpler rational structures has already been discussed in the literature~\cite{Abreu:2021fuk, Badger:2021nhg, Agarwal:2021grm,Abreu:2019odu,Chawdhry:2020for} and they have been utilized for expressing the amplitudes in a more compact from. On the other hand, it seems to us that the possible true structure behind these independent rational functions has not yet been fully explored and we hope to return to this in a future publication.

To aid future comparisons, benchmark evaluations of the finite remainders in one kinematic point are presented in appendix~\ref{sec:benchmarks}.

We have performed a number of checks on our results. When computing the 2-loop finite remainders for the two processes (\ref{eq:process}), we have verified that the poles in $\epsilon$ cancel. We have checked our tree-level and 1-loop results against the library {\tt Recola}~\cite{Actis:2016mpe}. We have also verified that the dependence on $\tr5$ (the one originating from the projector) drops out in the spin-averaged finite remainder. 

Recently, the spin-averaged amplitudes for the two processes (\ref{eq:process}) were calculated in ref.~\cite{Agarwal:2021grm}. We have found complete agreement for all terms except for those containing $\ep_5$. We have investigated the origin of this discrepancy. In the process of doing this we observed that we could reproduce the results of that reference if we do not flip the sign of $\ep_5$ under permutations during the mapping of masters to pentagon functions as well as during the numerical crossing of momenta in order to obtain all other helicities. As explained above such a treatment of $\ep_5$ is inconsistent. In order to verify this, we have calculated the terms of order $\epsilon$ and $\epsilon^2$ of the one-loop pentagon integral which are sensitive to the treatment of the parity-odd invariant and functions. We have verified that the calculation in terms of pentagon functions as described above agrees with a direct numerical calculation of this integral with the program {\tt pySecDec}~\cite{Borowka:2017idc}. Our interpretation of the above result is that the disagreement between our calculation and ref.~\cite{Agarwal:2021grm} is due to an inconsistent treatment in ref.~\cite{Agarwal:2021grm} of $\ep_5$ under permutations.

\centerline{\bf Note Added}

After this paper was completed but before it was submitted for publication we learned that the authors of ref.~\cite{Agarwal:2021grm} have independently discovered the inconsistency in the sign treatment of $\ep_5$ mentioned above. We now find complete agreement between our results and their corrected result, which should appear in an updated version of ref.~\cite{Agarwal:2021grm}.

\begin{table}[]
\centering
\begin{tabular}[t]{|l|c|l|c|}
\hline
$q\bar{q} \to g\g\g$ & \# tot./ \# ind. &$qg \to q\g\g$ & \# dep. / \# ind. \\
\hline
$\mathcal{R}^{+----,(2),Q_q^2,N_c^2}$ &   96 /  33 & $\mathcal{R}^{+-+--,(2),Q_q^2,N_c^2}$ & 6125 /  66 \\
$\mathcal{R}^{+----,(2),Q_q^2,n_f}$   &   48 /  22 & $\mathcal{R}^{+-+--,(2),Q_q^2,n_f}$   &   85 /  27 \\
$\mathcal{R}^{+----,(2),Q_{q'}^2,n_f}$&    6 /   2 & $\mathcal{R}^{+-+--,(2),Q_{q'}^2,n_f}$&   36 /   8 \\
$\mathcal{R}^{+-+--,(2),Q_q^2,N_c^2}$ & 7266 /  66 & $\mathcal{R}^{++++-,(2),Q_q^2,N_c^2}$ & 6200 / 101 \\
$\mathcal{R}^{+-+--,(2),Q_q^2,n_f}$   &  504 /  27 & $\mathcal{R}^{++++-,(2),Q_q^2,n_f}$   &  478 /  59 \\
$\mathcal{R}^{+-+--,(2),Q_{q'}^2,n_f}$&   58 /   8 & $\mathcal{R}^{++++-,(2),Q_{q'}^2,n_f}$&   50 /   8 \\
$\mathcal{R}^{+--+-,(2),Q_q^2,N_c^2}$ & 7252 / 101 & $\mathcal{R}^{+++--,(2),Q_q^2,N_c^2}$ &   92 /  33 \\
$\mathcal{R}^{+--+-,(2),Q_q^2,n_f}$   &  736 /  59 & $\mathcal{R}^{+++--,(2),Q_q^2,n_f}$   &   58 /  22 \\
$\mathcal{R}^{+--+-,(2),Q_{q'}^2,n_f}$&   58 /   8 & $\mathcal{R}^{+++--,(2),Q_{q'}^2,n_f}$&    4 /   2 \\
                                      &            & $\mathcal{R}^{+-++-,(2),Q_q^2,N_c^2}$ & 6216 / 101 \\
                                      &            & $\mathcal{R}^{+-++-,(2),Q_q^2,n_f}$   &  472 /  59 \\
                                      &            & $\mathcal{R}^{+-++-,(2),Q_{q'}^2,n_f}$&   50 /   8 \\
                                      &            & $\mathcal{R}^{+++++,(2),Q_q^2,N_c^2}$ & 6125 /  66 \\
                                      &            & $\mathcal{R}^{+++++,(2),Q_q^2,n_f}$   &   85 /  27 \\
                                      &            & $\mathcal{R}^{+++++,(2),Q_{q'}^2,n_f}$&   36 /   8 \\
                                      &            & $\mathcal{R}^{+-+++,(2),Q_q^2,N_c^2}$ &   92 /  33 \\
                                      &            & $\mathcal{R}^{+-+++,(2),Q_q^2,n_f}$   &   58 /  22 \\
                                      &            & $\mathcal{R}^{+-+++,(2),Q_{q'}^2,n_f}$&    4 /   2 \\
\hline

\end{tabular}
\caption{Total (tot.) vs. independent (ind.) number of rational functions of the finite remainders}
\label{table:rational}
\end{table}

\section{Conclusions}\label{sec:conclusion}

In this work we have calculated the finite remainders for the complete set of 2-loop QCD leading-colour helicity amplitudes for the processes $q\bar{q}\rightarrow g\gamma\gamma$ and $q g \rightarrow q \gamma \gamma$.
The results are obtained in a compact, fully-analytical form and can be found in the supplementary material attached to this paper.

These amplitudes enable the calculation of the double-virtual QCD corrections to $\gamma \gamma j$-production at hadron colliders. We expect that the calculation of this process in NNLO QCD is achievable using the framework used in our calculation of $\3g$-production at the LHC~\cite{Chawdhry:2019bji}.

In combination with the recently-calculated 3-loop QCD amplitude $q\bar{q} \rightarrow \gamma \gamma$~\cite{Caola:2020dfu}, our results open the door to the computation of the N\textsuperscript{3}LO QCD corrections to $\gamma \gamma$ production.

\begin{acknowledgments}
The work of M.C. was supported by the Deutsche Forschungsgemeinschaft under grant 396021762 - TRR 257. The research of A.M. and R.P. has received funding from the European Research Council (ERC) under the European Union's Horizon 2020 Research and Innovation Programme (grant agreement no. 683211). A.M. was also supported by the UK STFC grants ST/L002760/1 and ST/K004883/1. The research of H.C. is supported by the ERC Starting Grant 804394 HipQCD. A.M. acknowledges the use of the DiRAC Cumulus HPC facility under Grant No. PPSP226.
\end{acknowledgments}

\appendix

\section{Benchmark numerical results for the finite remainder}
\label{sec:benchmarks}

In this appendix we present numerical benchmark results for both amplitudes (\ref{eq:process}) evaluated at the following point
\begin{equation}
\mu=1,~~ x = \{1, 43/157, 83/157, 61/157, 37/157, i \sqrt{10196683}/24649\}\,,
\label{eq:point}
\end{equation}
where $\mu$ is the renormalization scale and $x$ is defined in eq.~(\ref{eq:x}). 

In table~\ref{tab:benchmark-sum} we give the values of the spin- and colour-summed squared two-loop finite reminders
$$\sum_{\bar h}\sum_{\rm color} 2{\rm Re}\left(\mathcal{F}^{(0)} \mathcal{F}^{(2)} \right)\,.$$
For each amplitude we show separately the coefficients of each one of the three charge/colour structures appearing in eqs.~(\ref{eq:F-R},\ref{eq:F-R-tilde}). The complete squared amplitude is then obtained by adding these three numbers each multiplied by its corresponding charge/color factor. We note that the squared amplitude is independent of the helicity conventions. 
\begin{table}[h]
\centering
\begin{tabular}[t]{|c|c|c|c|}

\hline
  &  $Q_q^2 N_c^2$ & $Q_q^2 n_f$ & $Q_{q'}^2 n_f$ \\
\hline
$q\bar{q} \to g \g\g$ & $156620.2$  & $-22398.6$ & $-1382.95$ \\
$qg \to q \g\g$ & $-1772.85$ & $-564.471$ & $4323.60$ \\
\hline
\end{tabular}
\caption{Benchmark evaluations of the squared two loop finite remainders in the point eq.~(\ref{eq:point}). Shown are the coefficients of the charge/color structures.}
\label{tab:benchmark-sum}
\end{table}

In table~\ref{tab:benchmark-F} we present the values of the coefficients $\mathcal{R}^{(2)}$ which are defined in eqs.~(\ref{eq:F-R},\ref{eq:F-R-tilde}). The reason we show these coefficients, instead of the complete two-loop finite remainders, is that the coefficients $\mathcal{R}$ are independent of the helicity conventions. Helicities whose values are not explicitly shown are zero. As in table~\ref{tab:benchmark-sum} above, the values are further split by the charge/color structures appearing in eqs.~(\ref{eq:F-R},\ref{eq:F-R-tilde}).
\begin{table}[h]
\centering
\begin{tabular}[t]{|c|c|c|c|}

\hline
Helicity & $\mathcal{R}^{(2),Q_q^2,N_c^2}$ & $\mathcal{R}^{(2),Q_q^2,n_f}$ & $\mathcal{R}^{(2),Q_{q'}^2,n_f}$ \\
\hline
\multicolumn{4}{c}{$q\bar{q} \to g \g\g$} \\
\hline
$+----$ & $ -9.0529 -2.1449i $ & $ 1.8348 +0.19391i $ & $ -0.0051350 -0.043221i $ \\
$-+---$ & $ -9.1772 +0.054019i $ & $ 1.8181 -0.22935i $ & $ 0.036676 +0.023436i $ \\
$-++++$ & $ -9.2761 -0.30316i $ & $ 1.8348 -0.19391i $ & $ -0.032907 -0.028487i $ \\
$+-+++$ & $ -8.9122 -2.2729i $ & $ 1.8181 +0.22935i $ & $ -0.0011607 +0.043509i $ \\
$+-+--$ & $ 29.541 +2.3640i $ & $ -3.3044 -0.36765i $ & $ -0.98915 -1.3505i $ \\
$-++--$ & $ 25.228 -3.0190i $ & $ -2.0349 +0.69140i $ & $ -0.67582 -1.5260i $ \\
$-+-++$ & $ 30.472 -1.4840i $ & $ -3.3044 +0.36765i $ & $ -0.56354 -1.5763i $ \\
$+--++$ & $ 24.168 +3.2096i $ & $ -2.0349 -0.69140i $ & $ -0.88457 -1.4153i $ \\
$+--+-$ & $ 27.867 +24.401i $ & $ -3.2863 -5.2198i $ & $ -0.38329 -0.51472i $ \\
$+---+$ & $ 131.03 -151.42i $ & $ -18.442 +33.892i $ & $ -0.71375 -2.1654i $ \\
$-+-+-$ & $ 76.984 +16.456i $ & $ -12.948 -1.6300i $ & $ -0.39323 -0.50343i $ \\
$-+--+$ & $ -53.799 -79.874i $ & $ 18.001 +12.817i $ & $ -1.3095 -1.9617i $ \\
$-++-+$ & $ 72.634 +18.963i $ & $ -12.119 -2.2083i $ & $ -0.21126 -0.60598i $ \\
$-+++-$ & $ -51.989 -89.280i $ & $ 17.474 +15.070i $ & $ -1.3927 -1.8052i $ \\
$+-+-+$ & $ 29.848 +27.439i $ & $ -3.9115 -5.8865i $ & $ -0.19635 -0.60788i $ \\
$+-++-$ & $ 121.36 -156.87i $ & $ -16.282 +34.746i $ & $ -0.88995 -2.1843i $ \\
\hline
\multicolumn{4}{c}{$qg \to q \g\g$} \\
\hline
$++++-$ & $ 0.55209 -21.589i $ & $ 0.61496 +6.6073i $ & $ 0.49135 -3.8124i $ \\
$+++-+$ & $ 271.57 +25.079i $ & $ -56.389 -8.8970i $ & $ 17.075 -24.896i $ \\
$----+$ & $ -7.5468 +7.5743i $ & $ 2.7324 +0.49795i $ & $ 0.16991 -2.5756i $ \\
$---+-$ & $ -64.889 +100.82i $ & $ 26.351 -23.482i $ & $ 39.726 -4.2734i $ \\
$+-++-$ & $ 4.7570 -72.345i $ & $ -8.0046 +16.377i $ & $ 6.4309 +7.8100i $ \\
$+-+-+$ & $ 1.7860 -64.731i $ & $ -2.5875 +16.314i $ & $ -1.1489 -1.8506i $ \\
$-+--+$ & $ 17.411 -140.01i $ & $ -9.4010 +29.890i $ & $ -1.1923 +20.567i $ \\
$-+-+-$ & $ 21.245 -133.79i $ & $ -12.116 +28.169i $ & $ -0.45528 -3.0363i $ \\
$+-+--$ & $ 7.5129 -120.03i $ & $ -8.7572 +26.290i $ & $ 1.3220 -2.1771i $ \\
$-+-++$ & $ 13.107 -92.756i $ & $ -7.0224 +20.685i $ & $ 1.3220 +2.1771i $ \\
$+++++$ & $ -7.3151 -11.657i $ & $ 2.6393 +3.4849i $ & $ 1.6828 +1.6226i $ \\
$-----$ & $ 2.2900 -38.856i $ & $ -0.39700 +9.5972i $ & $ 1.6828 -1.6226i $ \\
$+++--$ & $ 3.1373 +0.084647i $ & $ -0.57085 -0.18319i $ & $ -0.42571 +0.64703i $ \\
$---++$ & $ 2.4545 +0.89149i $ & $ -0.53179 -0.29430i $ & $ -0.42571 -0.64703i $ \\
$+-+++$ & $ -1.9218 +1.5696i $ & $ 0.53536 -0.44837i $ & $ 0.30067 -0.71377i $ \\
$-+---$ & $ -2.9417 -1.1496i $ & $ 0.39743 +0.27087i $ & $ 0.30067 +0.71377i $ \\
\hline
\end{tabular}
\caption{Benchmark evaluations of the two loop finite remainders in the point eq.~(\ref{eq:point}).}
\label{tab:benchmark-F}
\end{table}

\section{Renormalisation constants}
\label{sec:app-renorm}

We perform the renormalisation in the $\overline{\text{MS}}$ scheme with $n_f$ massless fermions.
The UV renormalisation constant $Z_{\alpha_s} = Z_g^2$ is given by:
\begin{eqnarray}
  Z_g &=& 1 + {1\over \ep} \biggl(\frac{\alpha_s}{4\pi}\biggr) \frac{4 n_f T_F-11 C_A}{6}
        + \biggl(\frac{\alpha_s}{4\pi}\biggr)^2 \left[
          {1\over \ep} \left(\frac{-17 C_A^2}{6} + \frac{5 C_A n_f T_F}{3}+C_F n_f T_F\right) \right. \nonumber\\
      & & \left. + {1\over \ep^2}\left(\frac{121 C_A^2}{384} - \frac{11 C_A n_f T_F}{48}
             +\frac{2 n_f^2 T_F^2}{48}\right)\right]\; + \mathcal{O}\left(\alpha_s^3 \right).
\end{eqnarray}
The IR renormalisation constant depends on the partonic process. Up to order ${\cal O}(\alpha_s^2)$ we find 
\begin{equation}
 \mathbf{Z}^{q\bar{q}} = 1 + \frac{\alpha_s}{4\pi}\biggl(
   \frac{\gamma_0}{2\ep^2} + \frac{\Gamma_0}{2\ep}\biggr) +\biggl(\frac{\alpha_s}{4\pi}\biggr)^2 \biggl(
     \frac{\gamma_0^2}{8\ep^4} + \frac{\gamma_0(\Gamma_0 - \frac{3}{2} b_0)}{4\ep^3} + \frac{\gamma_1 +\Gamma_0(\Gamma_0-2 b_0)}{8\ep^2} +\frac{\Gamma_1}{4\ep}\biggr)\;,
\end{equation} 
with
\begin{equation}
b_0 = \frac{11 C_A - 4 n_f T_F}{3}\,,
\end{equation}
and with the following anomalous dimensions
\begin{equation}
\gamma_0 = -2 (C_A + 2 C_F)\,,
\end{equation}
\begin{equation}
\gamma_1 = \frac{2}{9} (2 C_F + C_A)(C_A(-67+18 \zeta_2) + 20 n_f T_F)\,,
\end{equation}
\begin{equation}
\Gamma_0 = -(6 C_F + b_0) - 2 C_A (l_{\mu_{13}}+l_{\mu_{23}}) + 2 (C_A-2C_F)l_{\mu_{12}}\,,
\end{equation}
\begin{eqnarray}
\Gamma_1 &=& \frac{1}{54}\biggl(
               54 C_F^2(-3+24 \zeta_2)+C_A^2(-1384+198\zeta_2)+8C_F T_F n_f(92+54\zeta_2)\nonumber\\
           &&-2C_A\left(C_F(961+594 \zeta_2)+2 T_F n_f(-128+18\zeta_2)\right) + 108(C_A^2+26C_AC_F-24C_F^2)\zeta_3\nonumber\\
           && +12 \left(C_A(-67 + 18\zeta_2) + 20 n_f T_F\right)\left(C_A(l_{\mu_{13}}+l_{\mu_{23}}) -(C_A-2C_F)l_{\mu_{12}}\right) \biggr)\,.
\end{eqnarray}
In the above equations we have introduced the following notation: $l_{\mu_{12}} = \log (-\mu^2/s_{12})$, $l_{\mu_{23}} = \log (\mu^2/s_{23}) $, and $l_{\mu_{13}} = \log (\mu^2/s_{13})$.

The factor $\mathbf{Z}^{qg}$ is obtained from $\mathbf{Z}^{q\bar{q}}$ by replacing $l_{\mu_{12}} \to l_{\mu_{23}}$ and $l_{\mu_{23}} \to l_{\mu_{12}}$.

\end{document}